\documentclass[11pt, a4paper]{article}
\usepackage{amsmath}

\newtheorem{theorem}{Theorem}

\newcommand{\benumerate}{\begin{enumerate}}
\newcommand{\eenumerate}{\end{enumerate}}

\newcommand{\bitemize}{\begin{itemize}}

\newcommand{\eitemize}{\end{itemize}}
\newcommand{\ep}{\epsilon}

\begin{document}

\title{On the classification of scalar evolutionary integrable equations in $2+1$ dimensions }
\author{V.S. Novikov and E.V. Ferapontov}
    \date{}
    \maketitle
    \vspace{-7mm}
\begin{center}
Department of Mathematical Sciences \\ Loughborough University \\
Loughborough, Leicestershire LE11 3TU \\ United Kingdom \\[2ex]
e-mails: \\ {\texttt{V.Novikov@lboro.ac.uk}}\\
 { \texttt{E.V.Ferapontov@lboro.ac.uk}}\\
\end{center}

\bigskip

\begin{abstract}

We consider evolutionary equations of the form $u_t=F(u, w)$ where $w=D_x^{-1}D_yu$ is the  nonlocality,  and the right hand side $F$ is polynomial in the derivatives of $u$ and $w$.
The recent paper \cite{FMN} provides a complete list of integrable third order equations of this kind. Here we extend the classification to fifth order equations.
Besides the known examples of
Kadomtsev-Petviashvili (KP), Veselov-Novikov (VN) and  Harry Dym (HD) equations, as well as fifth order analogues and modifications thereof,
our list contains a number of equations which   are apparently new. We conjecture that our examples exhaust the list of scalar polynomial integrable equations with the  nonlocality $w$.
The classification procedure consists of two steps. First, we classify  quasilinear systems which may (potentially) occur as dispersionless limits of integrable scalar evolutionary equations. After that we reconstruct dispersive terms based on the requirement of the inheritance of hydrodynamic reductions of the dispersionless limit by the full dispersive equation.

\bigskip

\noindent MSC: 35L40, 35Q51, 35Q58, 37K10, 37K55.

\bigskip

Keywords: dispersionless equations, hydrodynamic reductions,
dispersive deformations, integrability.
\end{abstract}

\newpage

\section{Introduction}

The classification of integrable  $1+1$ dimensional scalar evolutionary  equations,
$$
u_t=F(u),
$$
 has  been (and still is) a subject of active research within the   soliton community. Here $u(x, t)$ is a scalar potential, and $F$ denotes a differential expression which depends on $x$-derivatives of $u$ up to some finite order. Although the general classification problem is still out of reach, quite a few  important results were obtained under various additional assumptions on $F$ (such as polynomiality, linearity in the highest derivative, etc).
We refer to the review article  \cite{Mik0} for a detailed discussion of the classification techniques involved, extensive lists of integrable equations within particularly interesting subclasses, and references.

In this paper we  apply the novel perturbative approach outlined in \cite{FerM, FMN} to a similar  problem in $2+1$ dimensions, the area where very few classification results are currently available. The main challenge of higher dimensions is the non-locality of scalar evolutionary integrable equations: the corresponding right hand side $F$ must contain nonlocal variables whose differential structure was clarified in  \cite{Mik1}. Here we consider equations of the form
\begin{equation}
u_t=F(u, w)
\label{1}
\end{equation}
where $u(x, y, t)$ is a scalar field and $w=D_x^{-1}D_yu$ is the simplest nonlocality (equivalently, $w$ can be introduced  via the relation $w_x=u_y$). We  assume that the right hand side $F$ is {\it polynomial} in  the $x$- and $y$-derivatives of
$u$ and $ w$, while the dependence on  $u$ and $w$ themselves is allowed to be arbitrary. The paper \cite{FMN} provides a complete list  of integrable third order equations of the form (\ref{1}),
\begin{equation}
\label{nonsym}
\begin{aligned}
u_t&=\varphi u_x+\psi u_y +\eta w_y+\epsilon(...)+\epsilon^2(...),
\end{aligned}
\end{equation}
where  $\varphi, \psi$ and $ \eta$ are functions of $u$ and $w$, while
the terms at  $\epsilon$ and $\epsilon^2$ are assumed to be
homogeneous differential polynomials of  the order two and  three
in the derivatives of $u$ and $w$ (one can show that all terms at $\epsilon$  have to vanish).
We use the following weighting scheme: $u$ and $w$ are assumed to have order zero, their derivatives    $u_x, u_y, w_x, w_y$ are of order one, the expressions $u_{xx}, u_{xy}, u_{yy}, w_{yy}, u_x^2, u_xu_y,  u_y^2, u_xw_y, u_yw_y, w_y^2$ are of order two, and so on.
Assuming that the dispersionless limit of the equation (\ref{nonsym}),
\begin{equation}
\label{dnonsym}
\begin{aligned}
u_t&=\varphi u_x+\psi u_y +\eta w_y, ~~~~~ w_x=u_y,
\end{aligned}
\end{equation}
is linearly nondegenerate (the property to be clarified in Sect. 2.2), and  satisfies the condition $\eta\ne 0$ (which is equivalent to the requirement that the dispersion relation of the system (\ref{dnonsym}) defines an irreducible conic), we have the following result:

\begin{theorem}\cite{FMN} Up to invertible transformations, the examples below provide a  complete   list of integrable third order equations (\ref{nonsym}) with $\eta \ne 0$ whose dispersionless limit is linearly
nondegenerate:
\begin{align*}
 &{ KP ~ equation}& \qquad &
 u_{t} = u u_{x} +w_y{{+ \epsilon^{2} u_{xxx}}}, &\\
&{ modified ~ KP ~ equation}& \qquad &
u_{t} =(w- u^2/2)  u_{x} +w_y {{+\epsilon^2  u_{xxx}}},&\\
&{Gardner ~ equation}& \qquad &
 u_{t} =(\beta w- \frac{\beta^2}{2}u^2+\delta u)  u_{x} +w_y {{+\epsilon^2  u_{xxx}}},&\\
&{  VN ~ equation}& \qquad &
u_{t} =(u w)_{y} {{+\epsilon^2  u_{yyy}}},&\\
&{ modified ~ VN ~ equation}& \qquad &
u_{t} =(u w)_y {{+\epsilon^2  \left( u_{yy}-\frac{3}{4}\frac{u_y^2}{u}\right)_y}},&\\
&{HD ~ equation }& \qquad &
u_{t} =-2w  u_{y} +u w_y {{-\frac{\epsilon^2}{u}\left(\frac{1}{u}\right)_{xxx}}},\\
&{ deformed ~ HD~ equation }& \qquad &
u_{t} =\frac{\delta}{u^3}u_x-2w  u_{y} +u w_y {{-\frac{\epsilon^2}{u}\left(\frac{1}{u}\right)_{xxx}}},&\\
&{ Equation \ E_1 }& \qquad &
u_t=(\beta w+\beta^2u^2)u_x-3\beta u u_y+w_y+\epsilon^2
[B^3(u)-\beta u_x B^2(u)],&\\
&{Equation \ E_2 }& \qquad &
u_t=\frac{4}{3}\beta^2 u^3u_x+(w-3\beta u^2) u_y+uw_y+\ep^2
[B^3(u)-\beta u_xB^2(u)],&
\end{align*}
here $B=\beta u D_x-D_y$, $\beta$=const.
\end{theorem}

The main result of this paper is a generalisation of the above classification to fifth order equations,
\begin{equation}
\label{nonsym5}
\begin{aligned}
u_t&=\varphi u_x+\psi u_y +\eta w_y+\epsilon (...)+\epsilon^2(...)+\epsilon^3(...)+\epsilon^4(...),
\end{aligned}
\end{equation}
where the terms at  $\epsilon^k$  are assumed to be
homogeneous differential polynomials of  the order  $k+1$ 
in the derivatives of $u$ and $w$, respectively. We also assume that the $\epsilon^4$ term depends on at least one of the possible fifth derivatives $u_{xxxxx},u_{xxxxy},\ldots$, i.e. the equation (\ref{nonsym5}) is of order 5.

\begin{theorem}  Up to invertible transformations,  the examples below provide a  complete   list of integrable fifth order equations  (\ref{nonsym5}) with $\eta \ne 0$ whose dispersionless limit is linearly
nondegenerate:

\begin{align*}
 &{ BKP ~ equation}& \qquad &
u_{t} = 5 (u^{2} + w) u_{x} + 5 u u_{y} - 5 w_{y} +5 \epsilon^{2} (
u u_{xxx} +  u_{xxy} +  u_xu_{xx}) + \epsilon^{4}
 u_{xxxxx},&\\
&{ CKP ~ equation}& \qquad &
u_{t} = 5 (u^{2} + w) u_{x} + 5 u u_{y} - 5 w_{y} +5 \epsilon^{2} (
u u_{xxx} +  w_{xxx} + \frac{5}{2} u_xu_{xx}) +
\epsilon^{4} u_{xxxxx},&\\
&{HD_5 ~ equation }& \qquad &
 u_{t} =15w  u_{y} -5u
w_y+5{\epsilon^2}\left[\frac{u_{xxy}}{u^2}-\frac{3}{u}\left(\frac{u_xu_y}{u^2}\right)_x\right]
-\frac{\epsilon^4}{2u^2}\left(\frac{1}{u^2}\right)_{xxxxx}, &\\
&{Equation ~ E_3}&\qquad &
u_{t}=4\gamma^2\frac{u_x}{u^5}+5(3w-\frac{\gamma}{u^2})u_y-5uw_y &\\
&{ }&\qquad &
+5\epsilon^2\left[
\frac{\gamma}{2u^2}\left(\frac{1}{u^2}\right)_{xxx}+\frac{u_{xxy}}{u^2}-\frac{3}{u}\left(\frac{u_xu_y}{u^2}\right)_x\right]-
\frac{\epsilon^4}{2u^2}\left(\frac{1}{u^2}\right)_{xxxxx}, &\\
&{Equation ~ E_4}&\qquad &
u_{t} =4\gamma^2\frac{u_x}{u^5}+5(3w-\frac{\gamma}{u^2})u_y-5uw_y  &\\
&{ }&\qquad &
+5\epsilon^2\left[
\frac{\gamma}{3u}\left(\frac{1}{u^3}\right)_{xxx}-\gamma\frac{u_x}{u^7}-\left(\frac{1}{u}\right)_{xxy}+
\left(\frac{u_xu_y}{u^3}\right)_x-\frac{u_y}{4u^4}\left(2uu_{xx}-3u_x^2\right)\right] &\\
&{ }&\qquad &
-\epsilon^4\left[\frac{1}{2u^2}\left(\frac{1}{u^2}\right)_{xxxxx}- \frac{15}{16}\left(\frac{(2uu_{xx}-3u_x^2)^2}{u^8}
\right)_x\right]. & \\
\end{align*}
\end{theorem}

\noindent We point out that the last two examples from Theorem 2 are apparently new. The equation $E_3$  can be viewed as a deformation of the fifth order Harry Dym equation $HD_5$: it reduces to $HD_5$ when $\gamma=0$.
Although each equation appearing in Theorems 1-2 gives rise to an integrable  hierarchy,  the corresponding higher flows will not belong to the class (\ref{1}): they will necessarily have a more complicated nonlocality. Preliminary calculations suggest that there  exist no seventh order equations of the form (\ref{1}). This leads to the following

\medskip

\noindent {\bf Conjecture} { \it Up to invertible transformations, Theorems 1-2 provide a complete  list of integrable evolutionary equations of the form (\ref{1}) which are  polynomial in the derivatives of $u$ and $w$.}

\medskip

\noindent {\bf Remark.} The assumption of  polynomiality is essential: there exist examples of integrable equations of the form (\ref{1}) where the right hand side $F$ is an infinite series in $\epsilon$. As an illustration, let us consider
 integrable differential-difference equations of the Toda lattice,
$$
v_{t} =  v  \triangle_{-}(w),  ~~~
 w_{x} = \triangle_{+}(v),
$$
where
$$
 \triangle_{-}(w)=\frac{w(y)-w(y- \epsilon)}{\epsilon}, ~~~  \triangle_{+}(v)=\frac{v(y+ \epsilon) - v(y)}{\epsilon}.
$$
Introducing the variable $u$ by the formula $\triangle_{+}(v)=u_y$, one  can rewrite the equations of the Toda lattice in such a way that the nonlocality $w$ will be of the required form,
$$
u_t=D_y^{-1}\triangle_{+}\left(\triangle_{+}^{-1}(u_y)\triangle_{-}(w)\right), ~~~ w_x=u_y.
$$
Expanding the first equation in  powers of $\epsilon$ one obtains an infinite series,
$$
u_t=uw_y+\frac{\epsilon^2}{12}(uw_{yy})_y+..., ~~~ w_x=u_y.
$$
Examples of this type will be outside the scope of this paper.

\medskip

The structure of the paper is as follows.  Following \cite{FMN},  in Sect. 2.1 we review  the classification of  integrable quasilinear systems of the form (\ref{dnonsym}). In Sect 2.2 we outline the general procedure  which, starting with an integrable dispersionless system, allows one to systematically reconstruct dispersive corrections. This procedure is applied in Sect. 2.3 to the case of fifth order equations (\ref{nonsym5}). For the reader's convenience, in Sect. 3 we present Lax pairs for all equations appearing in Theorems 1-2.

\section{Proof of Theorem 2}

The proof consists of two steps. In Sect. 2.1  we review the classification of  integrable quasilinear systems (\ref{dnonsym}) which may (potentially) occur as dispersionless limits of integrable soliton equations. In Sect. 2.2 we discuss the general procedure of the reconstruction of dispersive corrections based on the requirement of the inheritance of hydrodynamic reductions. This procedure is applied to fifth order equations in Sect. 2.3, leading to the proof of Theorem 2.

\subsection{Classification of integrable  dispersionless limits}

For a  system of the form (\ref{dnonsym}),
$$
u_t=\varphi u_x+\psi u_y +\eta w_y, ~~~
 w_x=u_y,
$$
the integrability conditions were obtained in \cite{FMN}
based on the results of  \cite{Fer5}. They constitute an involutive  system of second order
PDEs for the coefficients $\varphi,
\psi$ and $\eta$,
$$
\varphi _{uu}= -\frac{\varphi _w^2+\psi _u \varphi _w-2 \psi _w
\varphi _u}{\eta}, ~~~
\varphi _{uw}= \frac{\eta _w \varphi _u}{\eta}, ~~~
\varphi _{ww}= \frac{\eta _w \varphi _w}{\eta },
$$
$$
\psi _{uu}= \frac{-\varphi _w \psi _w+\psi _u \psi _w-2
\varphi _w \eta _u+2 \eta _w \varphi _u}{\eta }, ~~~
\psi _{uw}= \frac{\eta _w \psi _u}{\eta }, ~~~
\psi _{ww}= \frac{\eta _w \psi _w}{\eta },
$$
$$
\eta _{uu}= -\frac{\eta _w \left(\varphi _w-\psi _u\right)}{\eta}, ~~~
\eta _{uw}= \frac{\eta _w \eta _u}{\eta }, ~~~
\eta _{ww}=\frac{\eta _w^2}{\eta };
$$
we assume $\eta \ne 0$: this is equivalent to the requirement that
the dispersion relation of the system (\ref{dnonsym}) defines an
irreducible conic. The integrability conditions  are  straightforward to solve.
First of all, the  equations for $\eta$ imply that, modulo
translations and rescalings, one can set $\eta=1$, $\eta=u$ or $\eta=e^wh(u)$.
We will consider all three possibilities case-by-case below.
Notice  that $\varphi$ and $\psi$ are defined up to additive
constants which can always be set equal to zero via the Galilean
transformations of the initial system (\ref{dnonsym}). Moreover,
the integrability conditions are form-invariant under
transformations of the form
$$
\tilde \varphi=\varphi -s\psi+s^2\eta, ~~~ \tilde \psi=
\psi-2s\eta, ~~~ \tilde \eta = \eta, ~~~  \tilde u = u, ~~~ \tilde
w=w+su,  ~~~ s=const,
$$
which correspond to the following transformations
preserving the structure of system  (\ref{dnonsym}):
 $$
 \tilde x=x-sy, ~~~ \tilde y=y, ~~~ \tilde u = u, ~~~ \tilde w=w+su.
 $$
 All our classification results are formulated modulo this  equivalence.

\medskip

\noindent {\bf Case 1: $\eta=1$}. Then the remaining equations
imply $\psi=\alpha w+f(u),\  \varphi=\beta w+g(u)$, where $f$ and
$g$ satisfy the linear  ODEs
$$
f''=\alpha(f'-\beta), ~~~ g''=2\alpha g'-\beta f'-\beta^2.
$$
The subcase $\alpha =0$ leads to polynomial solutions of the form
\begin{equation}
\psi=\gamma u, ~~~~
 \varphi=\beta w-\frac{1}{2}\beta (\beta+ \gamma)u^2+\delta u.
\label{pol}
\end{equation}
Up to equivalence transformations, the case $\alpha \ne 0$ leads
to exponential solutions,
\begin{equation}
\psi= w+\beta e^{ u}, ~~~~
 \varphi=\alpha e^{2 u},
\label{exp}
\end{equation}
where $\alpha, \beta,  \gamma $ are arbitrary constants.

 \medskip

\noindent {\bf Case 2: $\eta=u$}. Then the remaining equations
imply $\psi=\alpha w+f(u),\  \varphi=\beta w+g(u)$, where $f$ and
$g$ satisfy the linear ODEs
$$
uf''=\alpha(f'-\beta)-2\beta, ~~~ ug''=2\alpha g'-\beta
f'-\beta^2.
$$
The case $\alpha \notin \{0,  -1, -1/2\}$ leads to power-like
solutions of the form
\begin{equation}
\psi=\alpha w+\gamma u^{\alpha +1}, ~~~~
 \varphi=\delta u^{2\alpha +1}.
\label{power}
\end{equation}
The subcase $\alpha =0$ leads  to logarithmic solutions,
\begin{equation}
\psi=-2\beta u\ln u -\beta   u, ~~~~
 \varphi=\beta w+\beta^2u\ln^2 u+\delta u.
\label{log1}
\end{equation}
The subcase $\alpha =-1$ gives
\begin{equation}
\psi =-w+\gamma \ln u, ~~~~ \varphi=\delta/u. \label{log2}
\end{equation}
Finally, the subcase $\alpha =-1/2$ gives
\begin{equation}
\psi =-\frac{1}{2}w+\gamma \sqrt u, ~~~~ \varphi=\delta \ln u.
\label{log3}
\end{equation}

\medskip

\noindent {\bf Case 3: $\eta=e^wh(u)$}. Then the remaining
equations imply $\psi=e^ wf(u),\  \varphi=e^ wg(u)$ where $f$, $g$
and $h$ satisfy the nonlinear system of ODEs,
$$
h''=f'-g, ~~~~ g''h=2fg'-gf'-g^2, ~~~~ f''h=2hg'-2gh'+ff'-fg.
$$
 Although
the structure of the general solution  is this system is quite complicated, one
can show that Case 3 cannot occur as the  dispersionless limit of an
integrable  soliton equation.

\subsection{Reconstruction of  dispersive corrections}

Given an integrable dispersionless system of the form (\ref{dnonsym}), one has to reconstruct
dispersive terms. This can be done by requiring that all
hydrodynamic reductions of the dipersionless system are inherited
by its dispersive counterpart \cite{FerM, FMN}. Following \cite{FMN}, we will  illustrate this procedure
with the example of  the  KP equation,
$$
u_{t} = u u_{x} +w_y+ \epsilon^{2} u_{xxx}, ~~~~ w_x=u_y.
$$
The dispersionless KP (dKP)  equation,
$$
u_{t} = u u_{x}+w_y, ~~~~ w_x=u_y,
$$
possesses one-phase solutions of the form  $u=R$, $w=w(R)$ where
the phase $R(x, y, t)$ satisfies a pair of Hopf-type equations
\begin{gather}
\label{R}
\begin{aligned}
R_{y} = \mu R_{x} , ~~~~ R_{t} =(\mu^{2} + R) R_{x};
\end{aligned}
\end{gather}
here $\mu(R)$ is an arbitrary function, and $w'=\mu$.
Equivalently, one can say that Eqs. (\ref{R}) constitute a
one-component hydrodynamic reduction of the dKP equation. Although
the dKP equation is known to possess infinitely many $N$-component
reductions for arbitrary $N$ \cite{Gibb94, GibTsa96, GibTsa99,
Kodama},  one-component reductions will be sufficient for our
purposes. The main observation of \cite{FerM} is that {\it all}
one-component reductions (\ref{R})  can be deformed into
reductions of the full KP equation by adding appropriate
dispersive terms which are {\it polynomial} in the $x$-derivatives
of $R$. Explicitly, one has the following formulae for the
deformed one-phase solutions,
\begin{equation}
u=R, ~~~ w=w(R)+\epsilon^{2}\left(\mu' R_{xx} +\frac{1}{2}(\mu''-
(\mu')^3) R_{x}^2 \right)+ O(\epsilon^{4}), \label{uw_Def}
\end{equation}
notice that one can always assume that $u$ remains undeformed
modulo the Miura group \cite{Dub1}. The deformed equations
(\ref{R}) take the form
\begin{gather}
\label{R_Def}
\begin{aligned}
R_{y} =& \mu R_{x} \\
& +\epsilon^{2} \left(\mu' R_{xx} +\frac{1}{2}
(\mu''- (\mu')^3) R_{x}^2 \right)_x + O(\epsilon^{4}),\\
R_{t} =& (\mu^{2} + R) R_{x} \\
&+\epsilon^{2} \left( (2\mu \mu'+1)R_{xx}+( \mu \mu''-\mu
(\mu')^3+(\mu')^2/2) R_{x}^2 \right)_x  + O(\epsilon^{4}),
\end{aligned}
\end{gather}
see  \cite{FerM}.
In other words, the KP equation can be `decoupled' into a pair of $(1+1)$-dimensional equations (\ref{R_Def}) in  infinitely many ways, indeed, $\mu(R)$ is an arbitrary function. The series in (\ref{uw_Def}) and (\ref{R_Def}) contain  even powers of $\epsilon$ only, and do not terminate in general.

Conversely, the requirement of the inheritance of all
one-component reductions allows one to reconstruct  dispersive
terms:  given the dKP equation, let us look for a third order
dispersive extension in the form
\begin{equation}
u_{t} = u u_{x}+w_y+\epsilon(...)+\epsilon^2(...), ~~~~ w_x=u_y,
\label{3.1}
\end{equation}
where the terms at $\epsilon$ and $\epsilon^2$ are homogeneous
differential polynomials in the derivatives of $u$
and $w$ of the order two and three, respectively.
We require that all one-component reductions (\ref{R}) can be
deformed accordingly, so that we have the following analogues of
Eqs. (\ref{uw_Def}) and (\ref{R_Def}),
\begin{equation}
u=R, ~~~ w=w(R)+\epsilon (...)+ \epsilon^{2}(...)+
O(\epsilon^{3}), \label{3.2}
\end{equation}
and
\begin{equation}
R_{y} = \mu R_{x} +\epsilon (...)+
\epsilon^2(...)+O(\epsilon^{3}), ~~~~ R_{t} =(\mu^{2} + R)
R_{x}+\epsilon (...)+\epsilon^2(...)+O(\epsilon^{3}), \label{3.3}
\end{equation}
respectively.  In Eqs. (\ref{3.2}) and (\ref{3.3}), dots denote
terms which are polynomial in the derivatives of $R$. Substituting
Eqs. (\ref{3.2}) into (\ref{3.1}), and using (\ref{3.3}) along
with the consistency conditions $R_{ty}=R_{yt}$, one arrives at a
complicated set of relations allowing one to uniquely reconstruct
dispersive terms in (\ref{3.1}): not surprisingly, we obtain that
all terms at $\epsilon $ vanish, while the terms at $\epsilon^2$
result in the familiar KP equation. Moreover, one only needs to
perform calculations up to the order $\epsilon^4$  to arrive at
this result. It is important to emphasise that the above procedure
is required to work for {\it arbitrary } $\mu$: whenever one
obtains a differential polynomial in $\mu$ which has to vanish due
to the  consistency conditions, all its coefficients have to be
set equal to zero independently. Another observation is that the
reconstruction procedure does not necessarily lead to a unique
dispersive extension like in the dKP case: one and the same
dispersionless system may possess essentially non-equivalent
dispersive extensions. In particular, VN and modified VN equations from Theorem 1, as well as BKP and CKP equations from Theorem 2 have coinciding dispersionless limits.

\medskip

Let us now turn to the general case of dispersionless equations of
the form (\ref{dnonsym}),
$$
\begin{aligned}
u_t&=\varphi u_x+\psi u_y +\eta w_y, ~~~~
 w_x=u_y.
 \end{aligned}
$$
The corresponding one-component reductions are of the form $u=R$, $w=w(R)$ where
$R(x, y, t)$ satisfies a pair of Hopf-type equations
\begin{gather*}
\begin{aligned}
R_{y} = \mu R_{x} , ~~~~ R_{t} =(\varphi +\psi \mu+ \eta \mu^{2})
R_{x};
\end{aligned}
\end{gather*}
here $\mu(R)$ is an arbitrary function, and $w'=\mu$. Let us seek a
third order dispersive deformation of  system (\ref{dnonsym}) in the
form
$$
\begin{aligned}
u_t&=\varphi u_x+\psi u_y +\eta w_y+\epsilon
(...)+\epsilon^{2}(...), ~~~~
 w_x=u_y,
 \end{aligned}
$$
and postulate that one-phase solutions can be deformed
accordingly,
$$
u=R, ~~~ w=w(R)+\epsilon(...)+\epsilon^{2}(...)+ O(\epsilon^{3}),
$$
where
$$
R_{y} = \mu R_{x} +\epsilon (...)+\epsilon^2(...)+O(\epsilon^{3}),
~~~~ R_{t} =(\varphi +\psi \mu+ \eta \mu^{2})
R_{x}+\epsilon(...)+\epsilon^2(...)+O(\epsilon^{3}).
$$
Proceeding as outlined above we reconstruct  dispersive
terms.

\noindent {\bf Remark.} We point out that the formulae for
dispersive deformations contain the expression
$$
\eta_w \mu^3+(\psi_w+\eta_u)\mu^2+(\varphi_w+\psi_u)\mu+\varphi_u
$$
in the denominator. Since $\mu$ is assumed to be arbitrary, this
expression is nonzero unless $\varphi, \psi, \eta$ satisfy the
relations
\begin{equation}
\eta_w=0, ~~~ \psi_w+\eta_u=0, ~~~ \varphi_w+\psi_u=0, ~~~
\varphi_u=0. \label{lindeg}
\end{equation}
These relations characterise the so-called {\it totally linearly
degenerate systems}. Dispersive  deformations of such systems  do not inherit  hydrodynamic
reductions, and  require a different
approach which is beyond the scope of this paper.

\subsection{Classification of fifth order equations}

In this Section we summarize the classification results  for
integrable fifth order  equations (\ref{nonsym5}),
$$
u_t=\varphi u_x+\psi u_y +\eta w_y+\epsilon (...)+\epsilon^2(...)+\epsilon^3(...)+\epsilon^4(...),
$$
which are obtained by adding dispersive terms to integrable
dispersionless candidates from Sect. 2.1 (one can show that all terms at $\epsilon$ and $\epsilon^3$ have to vanish).  Thus, we follow the classification of Sect. 2.1. We  concentrate on the case when the $\epsilon^4$-terms  contain at least one fifth order derivative of $u$ or $w$, and skip all cases leading to third order equations which were already classified in \cite{FMN}.

\medskip

\noindent {\bf Case 1:} We  verified that the exponential
solutions (\ref{exp}) do not survive, so that all non-trivial
examples come from the polynomial case (\ref{pol}),
$$
\eta=1, ~~~~ \psi=\gamma u, ~~~~
 \varphi=\beta w-\frac{1}{2}\beta (\beta+ \gamma)u^2+\delta u.
$$
A detailed  analysis of dispersive
deformations leads to the constraints $\gamma=\beta, \ \delta=0$. Modulo rescalings, this gives  BKP/CKP equations.

\medskip

\noindent {\bf Case 2:} One can prove that none of the logarithmic
cases (\ref{log1}), (\ref{log2})  and (\ref{log3})  survive, so
that all non-trivial examples come from the power case
(\ref{power}),
$$
\eta=u, ~~~~ \psi=\alpha w+\gamma u^{\alpha +1}, ~~~~
 \varphi=\delta u^{2\alpha +1}.
$$
The further analysis leads to the only possibility $\alpha =-3$. Modulo rescalings, the case $\delta=\gamma=0$ gives the $HD_5$ equation. The case of nonzero $\delta$ and $\gamma$ leads to the new equations $E_3$ and $E_4$.

\medskip

\noindent {\bf Case 3:} One can show that no examples
from this class possess  fifth order dispersive extensions.

\section{Lax pairs}

For the reader's convenience, in this section we bring together Lax pairs for all equations appearing in Theorems 1-2. We emphasise that our classification scheme does not assume the existence of a Lax pair: these come as the result of  direct calculations once the classification is completed. We refer to \cite{Kon4, Wang} for an alternative approach to the classification of integrable systems in $2+1$ dimensions based on postulating the structure of a Lax pair.

\subsection{Third order equations}

Since both KP and modified KP equations are particular cases of the Gardner equation, we will skip the first two examples.

\noindent The {\bf Gardner equation},
$$
 u_{t} =(\beta w- \frac{\beta^2}{2}u^2+\delta u)  u_{x} +w_y {{+\epsilon^2  u_{xxx}}},
 $$
possesses the Lax pair \cite{Kon4}
\begin{gather*}
\begin{aligned}
& {\ep}^2\psi_{xx}+ \frac{\ep}{\sqrt 3}( \psi_y-\beta u\psi_x)+\frac{\delta}{ 6}u \psi=0,  \\
&\ep \psi_t=4\ep^3 \psi_{yyy}-\sqrt 3 \beta \ep^2(2\psi_{xx}+u_x\psi_x)+\ep (\beta w+ \frac{\beta^2}{2}u^2 +\delta u)\psi_x+\ep \frac{\delta}{2} u_x-\frac{\beta \delta}{4\sqrt 3}u^2+\frac{\delta}{2\sqrt 3} w.
\end{aligned}
\end{gather*}
The {\bf  VN  equation},
$$
u_{t} =(u w)_{y} {{+\epsilon^2  u_{yyy}}},
$$
possesses the Lax pair \cite{VesNov,  Nizhnik}
\begin{gather*}
\begin{aligned}
& {\ep}^2\psi_{xy}+\frac{1}{ 3}u \psi=0,  \\
&\psi_t=\ep^2 \psi_{yyy}+w\psi_{y}.
\end{aligned}
\end{gather*}
The {\bf  modified VN  equation},
$$
u_{t} =(u w)_y {{+\epsilon^2  \left( u_{yy}-\frac{3}{4}\frac{u_y^2}{u}\right)_y}},
$$
possesses the Lax pair \cite{Bogdanov}
\begin{gather*}
\begin{aligned}
& {\ep}^2\psi_{xy}-{\ep}^2\frac{u_y}{2u}\psi_x+\frac{1}{ 3}u \psi=0,  \\
&\psi_t=\ep^2 \psi_{yyy}+w\psi_{y}+\frac{1}{2}w_y \psi.
\end{aligned}
\end{gather*}
The { \bf  HD equation},
$$
u_{t} =-2w  u_{y} +u w_y {{-\frac{\epsilon^2}{u}\left(\frac{1}{u}\right)_{xxx}}},
$$
possesses the Lax pair \cite{Kon4}
\begin{gather*}
\begin{aligned}
& \frac{\ep}{u^2}\psi_{xx}+\frac{1}{\sqrt 3} \psi_y=0,  \\
&\psi_t=\frac{4\ep^2}{u^3 } \psi_{xxx}+\left(\frac{2\sqrt 3 \ep w}{u^2}-\frac{6\ep^2
u_x}{u^4}\right)
\psi_{xx}.
\end{aligned}
\end{gather*}
The { \bf deformed  HD equation},
$$
u_{t} =\frac{\delta}{u^3}u_x-2w  u_{y} +u w_y {{-\frac{\epsilon^2}{u}\left(\frac{1}{u}\right)_{xxx}}},
$$
possesses the Lax pair \cite{FMN}
\begin{gather*}
\begin{aligned}
& \frac{\ep^2}{u^2}\psi_{xx}+\frac{\ep}{\sqrt 3} \psi_y+\frac{\delta}{4u^2}\psi=0,  \\
&\psi_t=\frac{4\ep^2}{u^3 } \psi_{xxx}+\left(\frac{2\sqrt 3 \ep w}{u^2}-\frac{6\ep^2
u_x}{u^4}\right)
\psi_{xx}+\frac{\delta}{u^3} \psi_x+\left(-\frac{3\delta
u_x}{2u^4}+\frac{\sqrt 3 \delta w}{2\ep u^2}\right).
\end{aligned}
\end{gather*}
The {\bf  Equation  E$_1$},
$$
u_t=(\beta w+\beta^2u^2)u_x-3\beta u u_y+w_y+\epsilon^2
[B^3(u)-\beta u_x B^2(u)],
$$
possesses the Lax pair \cite{FMN}
\begin{gather*}
\begin{aligned}
\epsilon^2\psi_{xy} &= \epsilon^2\beta u \psi_{xx}+\frac{1}{3} \psi,  \\
\psi_t &=\epsilon^2\beta^3 u^3\psi_{xxx} -
\epsilon^2\psi_{yyy}+3\epsilon^2\beta^2uu_y\psi_{xx}+\beta
w\psi_x.
\end{aligned}
\end{gather*}
The {\bf Equation  E$_2$},
$$
u_t=\frac{4}{3}\beta^2 u^3u_x+(w-3\beta u^2) u_y+uw_y+\ep^2
[B^3(u)-\beta u_xB^2(u)],
$$
possesses the Lax pair \cite{FMN}
\begin{gather*}
\begin{aligned}
\epsilon^2\psi_{xy} &=\epsilon^2\beta u \psi_{xx}+\frac{1}{3}u\psi ,  \\
\psi_t &=\epsilon^2\beta^3u^3\psi_{xxx}
-\epsilon^2\psi_{yyy}+3\epsilon^2\beta^2uu_y\psi_{xx}+\frac{\beta^2}{3}u^3\psi_x+w\psi_y+\beta uu_y\psi.
\end{aligned}
\end{gather*}

\subsection{Fifth order equations}

The {\bf BKP equation},
\begin{align*}
u_{t} = 5 (u^{2} + w) u_{x} + 5 u u_{y} - 5 w_{y} +5 \epsilon^{2} (
u u_{xxx} +  u_{xxy} +  u_xu_{xx}) + \epsilon^{4}
 u_{xxxxx},&\\
\end{align*}
possesses the Lax pair  \cite{Kon4}
\begin{gather}
\begin{align*}
&\psi_{y} +u \psi_{x}+\epsilon^2 \psi_{xxx}=0,  \\
&\psi_t +5(u^2-w)\psi_x+\epsilon^2(15u\psi_{xxx}+15u_x\psi_{xx}+10u_{xx}\psi_x)+9\epsilon^4\psi_{xxxxx}=0.
\end{align*}
\end{gather}
The {\bf CKP equation},
\begin{align*}
u_{t} = 5 (u^{2} + w) u_{x} + 5 u u_{y} - 5 w_{y} +5 \epsilon^{2} (
u u_{xxx} +  u_{xxy} + \frac{5}{2} u_xu_{xx}) + \epsilon^{4}
 u_{xxxxx},&\\
\end{align*}
possesses the Lax pair  \cite{Kon4}
\begin{gather}
\begin{align*}
&\psi_{y} +u \psi_{x}+\frac{1}{2}u_x\psi +\epsilon^2 \psi_{xxx}=0,  \\
&\psi_t +5(u^2-w)\psi_x+5(uu_x-\frac{1}{2}u_y)\psi+
\epsilon^2(15u\psi_{xxx}+\frac{45}{2}u_x\psi_{xx}+\frac{35}{2}u_{xx}\psi_x+5u_{xxx}\psi)+9\epsilon^4\psi_{xxxxx}=0.
\end{align*}
\end{gather}
The {\bf HD equation},
\begin{align*}
 u_{t} =15w  u_{y} -5u
w_y+5{\epsilon^2}\left[\frac{u_{xxy}}{u^2}-\frac{3}{u}\left(\frac{u_xu_y}{u^2}\right)_x\right]
-\frac{\epsilon^4}{2u^2}\left(\frac{1}{u^2}\right)_{xxxxx},\\
\end{align*}
possesses the Lax pair  \cite{Kon4}
\begin{gather}
\begin{align*}
&\psi_{y} +\frac{\epsilon^2}{u^3} \psi_{xxx}=0,  \\
&\psi_t +15\epsilon^2\frac{w}{u^3}\psi_{xxx}+\epsilon^4\left[\frac{9}{u^5}\psi_{xxxxx}-45\frac{u_x}{u^6}\psi_{xxxx}+\frac{15}{u^3}\left(\frac{1}{u^2}\right)_{xx}\psi_{xxx}\right]=0.
\end{align*}
\end{gather}
The {\bf  Equation  E$_3$},
\begin{gather*}
 \begin{aligned}
u_{t}=&4\gamma^2\frac{u_x}{u^5}+5(3w-\frac{\gamma}{u^2})u_y-5uw_y\\
&+5\epsilon^2\left[
\frac{\gamma}{2u^2}\left(\frac{1}{u^2}\right)_{xxx}+\frac{u_{xxy}}{u^2}-\frac{3}{u}\left(\frac{u_xu_y}{u^2}\right)_x\right]-
\frac{\epsilon^4}{2u^2}\left(\frac{1}{u^2}\right)_{xxxxx},\\
\end{aligned}
\end{gather*}
possesses the Lax pair
\begin{eqnarray*}
&&\psi_y-\frac{\gamma}{u^3}\psi_x+\frac{\epsilon^2}{u^3}\psi_{xxx}=0,\\
&&\psi_t+\left(\frac{6\gamma^2}{u^5}-\frac{15\gamma w}{u^3} \right)\psi_x+15\epsilon^2\left[ \left(\frac{w}{u^3}-\frac{\gamma}{u^5}\right)\psi_{xxx}+\frac{3\gamma u_x}{u^6}\psi_{xx}+\frac{2\gamma}{u^3}\left(\frac{u_x}{u}\right)_x\psi_x  \right]\\
&&+\epsilon^4\left[\frac{9}{u^5}\psi_{xxxxx}-45\frac{u_x}{u^6}\psi_{xxxx}+\frac{15}{u^3}\left(\frac{1}{u^2}\right)_{xx}\psi_{xxx}\right].
\end{eqnarray*}
The {\bf Equation   E$_4$},
\begin{gather*}
\begin{aligned}
u_{t} =&4\gamma^2\frac{u_x}{u^5}+5(3w-\frac{\gamma}{u^2})u_y-5uw_y \\
&+5\epsilon^2\left[
\frac{\gamma}{3u}\left(\frac{1}{u^3}\right)_{xxx}-\gamma\frac{u_x^3}{u^7}-\left(\frac{1}{u}\right)_{xxy}+
\left(\frac{u_xu_y}{u^3}\right)_x-\frac{u_y}{4u^4}\left(2uu_{xx}-3u_x^2\right)\right]\\
&-\epsilon^4\left[\frac{1}{2u^2}\left(\frac{1}{u^2}\right)_{xxxxx}- \frac{15}{16}\left(\frac{(2uu_{xx}-3u_x^2)^2}{u^8}
\right)_x\right],
\end{aligned}
\end{gather*}
possesses the Lax pair
\begin{eqnarray*}
&& \psi_y+\left(\frac{\gamma
u_x}{u^4}+\frac{u_y}{2u}\right)\psi-\frac{\gamma}{u^3}\psi_x+\epsilon^2\left[\frac{1}{u^3}\psi_{xxx}+
\left(\frac{3}{2}\frac{u_{xx}}{u^4}-\frac{9}{4}\frac{u_x^2}{u^5}\right)\psi_x+\left(\frac{1}{2u}\left(\frac{u_{xx}}{u^3}\right)_x+
\frac{3}{4}\frac{u_x^3}{u^6}\right)\psi\right]=0,
\end{eqnarray*}
\begin{eqnarray*}
&&\psi_t+\left(\frac{15(u^3u_y+2\gamma u_x)w}{2u^4}-\frac{10\gamma u_y}{u^3}-\frac{10\gamma^2u_x}{u^6}-\frac{5}{2}w_y\right)\psi+\left(6\frac{\gamma^2}{u^5}
-15\frac{\gamma w}{u^3}\right)\psi_x+\\
&&+\epsilon^2\left[\frac{15(wu^2-\gamma)}{u^5}\psi_{xxx}+\left(\frac{60\gamma u_x}{u^6}+\frac{15u_y}{2u^3}\right)\psi_{xx}+\right.\\ &&\left.+\left(\frac{30\gamma}{u^2}\left(\frac{u_x}{u^4}\right)_x +\frac{15u_xu_y}{2u^4}+\frac{(90uu_{xx}-135u_x^2)w}{4u^5}\right)\psi_x+\right.\\
&&\left.+\left(5\gamma\left(\frac{u_{xx}}{u^6}\right)_x-\frac{15\gamma}{u}\left(\frac{u_x^2}{u^6}\right)_x+\frac{5}{2}\left(\frac{u_{xy}}{u^3}\right)_x+\frac{15}{2u}w\left(\frac{u_{xx}}{u^3}\right)_x+\frac{45u_x^3}{4u^6}w+\frac{15u_x^2u_y}{2u^5} \right)\psi \right]+\\
&&+\epsilon^4\left[\frac{9}{u^5}\psi_{xxxxx}-\frac{45u_x}{u^6}\psi_{xxxx}+\left(-\frac{15u_{xx}}{2u^6}+\frac{225u_x^2}{4u^7}\right)\psi_{xxx}+
\left(30\left(\frac{u_{xx}}{u^6}\right)_x+180\frac{u_x^3}{u^8}\right)\psi_{xx}+\right.\\
&&\left.+\left(\frac{45}{2}\left(\frac{u_{xxx}}{u^6}\right)_x-60\left(\frac{u_xu_{xx}}{u^7}\right)_x-\frac{105u_{xx}^2}{2u^7}+\frac{825}{6u^4}\left(\frac{u_x^3}{u^4}\right)_x-\frac{235u_x^4}{8u^9} \right)\psi_x\right.\\&&\left.+\left(-\left(\frac{1}{u^5}\right)_{xxxxx}+\frac{195}{2}\left(\frac{u_xu_{xxx}}{u^7}\right)_x+\frac{135}{2}\left(\frac{u_{xx}^2}{u^7}\right)_x-\frac{4605}{4}\left(\frac{u_x^2u_{xx}}{u^8}\right)_x-\frac{165u_xu_{xx}^2 }{2u^8}+\right.\right.\\
&&\left.\left.+\frac{3375}{2}\left(\frac{u_x^4}{u^9}\right)_x+\frac{3645u_x^5}{8u^{10}}   \right)\psi \right].
\end{eqnarray*}
We do not exclude a possibility that  simpler Lax pair can be found in this case.


\section*{Acknowledgements}

The research of EVF was partially supported  by the European Research Council Advanced Grant  FroM-PDE.

\end{document}